# Deciphering Pluto's Haze: How a Solar-Powered Vapor-Pressure Plume Shapes Its Bimodal Particle Size Distribution

Sihe Chen[1], Danica Adams[2], Siteng Fan[3], Peter Gao[4], Eliot Young[5], and Yuk Yung[1]

1 Department of Geological and Planetary Sciences, California Institute of Technology, Pasadena, CA 91125, United States of America
2 Earth & Planetary Sciences Department, Harvard University, Cambridge, MA 02138, United States of America
3 College of Science, Southern University of Science and Technology, Shenzhen, Guangdong 518055, China
4 Earth & Planets Laboratory, Carnegie Institute of Science, Washington, DC 20015, United States of America
5 Space Studies Department, Southwest Research Institute, Boulder, CO 80302, United States of America

**Abstract**

Combining findings from New Horizons' suite of instruments reveals a bimodal haze particle distribution within Pluto's atmosphere, which haze models have not been able to reproduce. We employ the photochemical and microphysics KINAERO model to simulate seasonal cycles and their impact on the haze distribution. We find that the smaller spherical particle mode can be generated through photochemistry and coagulation, while the larger aggregate mode are formed by surface volatile deposits sublimating and subsequently lofting such particles upwards. This process leads to two critical insights: firstly, only aggregates composed of small monomers can be lofted, and secondly, this aggregate distribution is transient and tied closely to Pluto's seasonal cycles. We thus anticipate that the bimodal distribution will vanish in the coming decades as Pluto's atmosphere undergoes collapse.

**Introduction**

Located in the Kuiper belt, Pluto has a cold and tenuous atmosphere. However, the highly complex organic photochemistry within, initiated by methane photodissociation, resembles the photochemical processes of the Archean Earth, where organics also originated from methane (Kasting et al., 1983). Hence, studying Pluto helps us understand how organics were first synthesized on Earth, which further leads to understanding the origin of life. Such processes are important for studying the astrobiology of exoplanets with reducing atmospheres as well (e.g. Adams et al., 2022).

Comparing photochemical models of Pluto with observations provided us with insights to the composition of the atmosphere. Krasnopolsky & Cruikshank (1999) modelled the photochemistry of Pluto's atmosphere based on stellar occultation measurements and concluded that the photochemistry of Pluto resembles that of Titan more than that of Triton. With the *New Horizons* flyby, both the understanding of Pluto's atmospheric structure and photochemistry were improved by Wong et al., (2017), who found a surface $CH_4$ mixing ratio of 0.004, a constant-with-altitude eddy diffusivity of $10^3$ $cm^2s^{-1}$, and evidence of condensation of hydrocarbon and nitrile ices on haze particles.

The haze of Pluto is one of the most important observations from New Horizons (Gladstone et al., 2016) and has a profound impact on the thermal structure of Pluto's atmosphere. Due to their larger solar heating and thermal cooling rates, haze particles dominate the radiative balance from the surface to an altitude of 700 km (Zhang et al., 2017). The haze also contributes to the surface material on Pluto and potentially interacts with the ices at the surface to form variations in Pluto's surface colors (Grundy et al., 2018).

With the *New Horizons* flyby, our understanding of Pluto's haze has greatly improved (Gladstone et al., 2016; Stern et al., 2015; Young et al., 2018, Cheng et al. 2017). For example, Gao et al. (2017) showed using microphysical modeling that the haze UV extinction can be explained by tholin-like fractal aggregates produced from photochemistry and coagulation, with the particle size and number density controlled by the particle charge. Specifically, for a particle charge to radius ratio of 30 $e^-/\mu m$, aggregates grew to ~0.1 µm near the surface, consistent with estimates made at the time from optical scattered light measurements (Gladstone et al., 2016). They further showed that spherical particles underestimated UV extinction by a factor of 2 to 3 for the same haze production rate as the aggregate case and resulted in particle radii of a few tens of nm near the surface. However, a more comprehensive analysis of New Horizons observations of UV extinction and optical and near-infrared scattering made by Fan et al. (2022), in which four distinct observational datasets were integrated, established that the near-surface haze size distribution is actually bimodal, with a large (~1 µm) aggregate mode and a smaller (<0.1 µm) spherical particle mode. While tuning the particle charge to radius ratio in the model of Gao et al. (2017) can produce either of these two modes, producing both simultaneously near the surface was not possible, and no other microphysical model has predicted two modes there even when ice condensation is considered (Stern et al. 2017; Lavvas et al. 2021). The origin of this bimodality thus remains unresolved, prompting further investigations into the atmospheric and haze processes of Pluto.

In this paper, we evaluate for the first time whether atmospheric collapse and replenishment can generate the near-surface bimodal haze distribution. With Pluto's eccentric orbit, its atmosphere undergoes significant depletion as it recedes from the sun, only to be replenished upon nearing its perihelion (Johnson et al., 2021). Sputnik Planitia plays a critical role in this process, acting as both a primary source and a sink for atmospheric gases, thereby significantly impacting atmospheric circulation (Bertrand et al., 2020). Our study introduces a comprehensive model that captures the photochemical processes throughout Pluto's yearly cycles, detailing the atmospheric collapse and subsequent regeneration. We then explore the microphysical and dynamical processes that give rise to the bimodal distribution of atmospheric particles, linking these phenomena to the atmospheric cycles dictated by Pluto's orbit and the influence of Sputnik Planitia.

**Model Description**

We develop a Pluto photochemical-microphysical model based on the Caltech/JPL photochemical model, KINETICS (Allen et al., 1981; Wong et al., 2017). In this model, we use simplified chemistry that contains only up to C2 chemistry for higher computational efficiency, as we aim to study the distribution of haze instead of solving for more complex chemistry.

Our model meticulously simulates the cyclical collapse and regeneration of Pluto's atmosphere over its annual cycle. Initially, the model achieves an equilibrium state using Pluto's current surface temperatures and thermal conditions, akin to the set up detailed by Wong et al. (2017). Following this, it tracks atmospheric pressure and temperature profile changes as outlined by Bertrand & Forget (2016) over a Pluto orbit. Among various subsurface thermal inertia (TI) values explored, we opted for a TI of 800 SI, though our findings are robust across different TI settings. As the model operates on a height coordinate system, it dynamically adjusts the vertical distribution of particle number density during atmospheric collapse, ensuring its pressure coordinate profile remains consistent. This adjustment reflects the surface condensation and subsequent removal of atmospheric gases from the lower atmosphere. Conversely, during atmospheric regeneration, these lowest layers are replenished with gas freshly sublimated from the surface. Given Pluto's frigid surface conditions, most molecular species, particularly higher-order hydrocarbons, condense permanently and do not return to the atmosphere during atmospheric regeneration. Therefore, we limit re-entrance into the atmosphere to only $N_2$ and $CH_4$, with $H_2$ never undergoing condensation.

In addition to the photochemistry, we incorporate the growth of spherical and aggregate aerosol particles into the model, with an aerosol mass production rate expressed as the production of small, initial seed particles equal to the destruction rate of methane, as in Gao et al. (2017). The coagulation of the aerosols is incorporated into the KINETICS model as chemical reactions, in a new model named "KINAERO". As in Gao et al. (2017) and Lavvas et al. (2010), the coagulation rate "reaction coefficient", between two bins $i, j$ is calculated as

$$\beta_{i,j} = 4\pi(D_i + D_j)(r_i + r_j)\alpha \quad (1)$$

Where

$$\alpha^{-1} = \frac{r_i + r_j}{r_i + r_j + \sqrt{\delta_i^2 + \delta_j^2}} + \frac{4(D_i + D_j)}{(r_i + r_j)\sqrt{\bar{v}_i^2 + \bar{v}_j^2}} \quad (2)$$

$$\delta_i = \frac{(2r_i + \lambda_i)^3 - (4r_i^2 + \lambda_i^2)^{\frac{3}{2}}}{6r_i\lambda_i} - 2r_i \quad (3)$$

Here $r_i$ is the radius of the particle in the $i$-th bin. The thermal velocity $\bar{v}$, mean free path $\lambda$, and particle diffusivity $D$ are found for bin number $i$ as

$$\bar{v}_i = \sqrt{\frac{8k_B T}{\pi M_i}} \quad (4)$$

$$\lambda_i = \frac{8D_i}{\pi \bar{v}_i} \quad (5)$$

$$D_i = \frac{k_B T f_{slip}}{6\pi \eta r_i} \quad (6)$$

where $k_B$ is the Boltzmann constant, $T$ is the temperature, $\eta$ is the dynamic viscosity of the gas, and $M_i$ the particle mass of the $i$-th bin. The slip correction factor is calculated as

$$f_{slip} = 1 + 1.257 Kn + 0.4 Kn \exp(-1.1 Kn^{-1}) \quad (7)$$

Here Kn is the Knudsen number. On Pluto, it can be shown that the Knudsen number is large:

$$Kn = \frac{\lambda_a}{r_i} = \frac{k_B T}{4\sqrt{2} r_{N_2}^2 r_i} \frac{T}{P} \gg 1 \quad (8)$$

Here $\lambda_a$ is the mean free path, $P$ is the atmospheric pressure, and $r_{N_2}$ is the hard-shell diameter of the nitrogen molecule. In this case, an approximation can be made such that

$$\delta_i \approx \lambda_i \quad (9)$$

Thus, the reaction coefficient can be written as

$$\beta_{ij} = \pi r_0^2 \left(N_i^{\frac{2}{D_f}} + N_j^{\frac{2}{D_f}}\right)\sqrt{\frac{k_B T}{\mu}} \quad (10)$$

Here $r_0$ is the initial seed particle size, $N_i$ the number of seed particles that comprise the coagulated particle in the $i$-th bin, and $D_f$ the fractal dimension ($D_f = 3$ for spheres). In the case of aggregates with $D_f < 3$, $r_0$ is the monomer size and $N_i$ is the number of monomers in an aggregate in the $i$-th bin. With this approximation, coagulation is described as chemical reactions with their reaction rates proportional to the square root of the current temperature. The yield ratio of the products is calculated so that the aerosol particles produced from coagulation have the same mass as the reactant aerosol particle to ensure mass conservation.

As photochemistry and coagulation can only produce one particle mode (Gao et al. 2017), we must look for an additional process to generate the observed bimodal haze size distribution. Here we posit that haze particles that have already sedimented onto the surface may be lofted back up into the atmosphere during atmospheric regeneration. When Pluto has gone through its aphelion and gets warmed up by the sun again, the "ground" that the particles rest on sublimates and becomes gaseous. This is, in general, a poorly understood process, but the condition for the particles to stay in the atmosphere is straightforward: if the sedimentation velocity of the particles is slower than the speed at which the sublimating gases are moving upwards into the atmosphere, then the particles can get lofted.

The lofting process enables an easier entrance of the haze particles into the atmosphere compared to dust entrainment by high-speed winds, which could be inhibited by an adhesion force between the surface matrix and the aerosol particles that vanishes when the surface itself is turned into the gaseous phase. We provide a conceptual understanding by determining the characteristic velocities for atmospheric accumulation and particle sedimentation. The atmosphere expands from minimal pressure to 2 Pa over approximately 50 years, with the most rapid phase achieving a 1 Pa increase within a decade, as detailed by Bertrand & Forget (2016). The surface upward wind speed is then calculated as

$$v_s = \frac{\dot{P}_s R_g T_s}{g P_s} \quad (11)$$

Where $P_s$ is the surface pressure and $\dot{P}_s$ is its rate of change, $R_g$ is the specific gas constant, $T_s$ is the surface temperature, and $g$ is the gravitational acceleration. The sedimentation velocity of the particles near the surface follows from Gao et al. (2017),

$$w_{sed} \approx 0.5 \frac{\rho_P g r_P}{\rho_g} \sqrt{\frac{\pi}{2 R_g T_s}}$$

$$= 0.5 \frac{\rho_P g r_P}{P_s} \sqrt{\frac{1}{2} \pi R_g T_s} \quad (12)$$

Where $\rho_P$ is the particle density, $r_P$ is the particle radius, and $\rho_g$ is the density of the gas. For the haze particles to get lofted, we need the ratio of the surface velocity to the sedimentation velocity to be larger than 1:

$$\frac{v_s}{w_{sed}} = \frac{\dot{P}_s}{\rho_p r_p g^2} \sqrt{\frac{8 R_g T_s}{\pi}} > 1 \quad (13)$$

For an aerosol particle with fractal dimension $D_f$ coagulated from $N$ initial particles (monomers, in the case of aggregates), $\rho_p = \rho_0 N^{1-\frac{3}{D_f}}$, where $\rho_0$ is the density of the aerosol material. Similarly, $r_P = r_0 N^{\frac{1}{D_f}}$. By inserting the constants, $g = 0.62\ m\ s^{-2}, R_g \approx R_{N_2} = 297\ J\ kg^{-1} K^{-1}, \rho_p = 800\ kg\ m^{-3}$, and assuming a constant value of $T_s = 40\ K$ for simplicity, we obtain

$$\frac{v_s}{w_{sed}} \approx 0.6 \frac{\dot{P}_s}{r_0 N^{1-\frac{2}{D_f}}} \quad (14)$$

From the equations above, we see that whether a particle can be lofted depends on the initial particle/monomer size, fractal dimension, and the rate of pressure increase. For aggregates with a fractal dimension of $D_f = 2$, which is what we consider here, the dependency on the number of monomers in the aggregate is removed. Moreover, if $D_f < 2$, more massive aggregates are selectively lofted since $N$ is raised to a negative power; $D_f < 2$ is suggested in Kutsop et al. (2021) and is within the uncertainties of Fan et al. (2022). In contrast, spherical particles ($D_f = 3$) leads to the opposite behavior where smaller particles are preferentially lofted. As our proposed lofting process is most efficient for large aggregates, we use it to model the observed large aggregate mode.

In the derivation, the rate of increase for the surface pressure, $\dot{P}_S$, is used. However, most of the sublimation flow comes from Sputnik Planitia, and induces a higher vertical speed than currently calculated, as well as a diverging/converging flow during sublimation/condensation. In our 1-dimensional model, we use vertical eddy diffusivity $k_{zz}$ to model the mixing induced by such flow. Additionally, the daily fluctuation of the surface pressure is ignored, as we do not expect it to have significant impact on the overall vertical profile of the aerosol particle distributions.

We also note that, for a particle at an arbitrary height $z$, the requirement for the particle to continue to ascend in the atmosphere has the same form

$$\frac{v}{w_{sed}} = \frac{\dot{P}(z)}{\rho_p r_p g^2(z)} \sqrt{\frac{8 R_g T(z)}{\pi}} > 1 \quad (15)$$

From the above equation, it is noted that the particles are better lifted where the atmosphere is warm. Due to the existence of a temperature inversion, the ascent speed of the haze particles does not decay as fast as $\dot{P}(z)$. This enables the particles to reach higher in the atmosphere when a temperature inversion is present.

## Results

We run the photochemical model for three Pluto years. While one Pluto year is 248 Earth years, the nearly full collapse of the atmosphere every Pluto winter provides fast convergence to the model, i.e. the atmosphere has very poor long-term memory. In Figure 1, the distribution of the C2 species over time and altitudes are plotted as contours. The 248-year cycles of a Pluto year are visible, and the patterns for all three species are alike. The number density profile agrees with the New Horizon observations above 100 km, but the depletion of C2 species in the lower atmosphere, caused by freshly sublimated $N_2$ and $CH_4$ gas pushing the C2 species upwards, is not observed, particularly below 50 km (Young et al., 2018). This is possibly due to mixing induced by the converging/diverging flow in the lower atmosphere at Sputnik Planitia, which our 1D model cannot capture, or C2 species reentering the atmosphere, either in gaseous form, or in particle form and then evaporating at higher altitudes. Figure 2 shows that the photochemical destruction rate of methane (and thus the production rate of aerosols) has a particularly strong peak when the atmosphere is the most massive (at years 0-25 and 225-250 since 2020). This peak does not come from only the increase in methane concentration, but it is also related to the increase in solar flux at perihelion.

The distribution of haze produced from photochemistry and coagulation is shown in Figure 3. Here we have assumed spherical haze particles ($D_f = 3$), as we hypothesize that lofting provides the larger aggregate mode. In this simulation, the first particle bin (the "initial seed particles" previously mentioned) corresponds to a radius of 0.04 μm, and the $i$-th bin is formed from the coagulation of $N_i = 4^{i-1}$ first-bin particles. The computed near-surface particle number density and size distribution is in good agreement with the spherical haze particle scenario from Gao et al. (2017), though the current simulation has a coarser bin resolution. This shows that the smaller, spherical mode revealed by Fan et al. (2022) can be formed via photochemistry and coagulation if the haze particles remain compact.

To simulate the lofted aggregate mode, which is not coupled to the photochemistry, we use a simple scheme to compute the concentration of the aerosol particles that enter the atmosphere together with the gas. Since the lofting process mostly only affects the lower atmosphere, we divide the lower 40 km of the atmosphere into 50 layers in the simulation. The aerosol flux is proportional to the gas flux during atmosphere growth and is turned off during atmosphere collapse. We parameterize the surface flux of the aggregates using a constant number density and we find that a value of 0.139 cm$^{-3}$ produces an aerosol distribution that matches the observation, as shown in Figure 4. Our results also feature the observed inversion of concentration at about 20 km altitude. An additional inversion below 10 km is predicted, but this is out of the retrieval range reported in Fan et al. (2022). These inversions are generated by the enhanced lofting ability at higher temperatures, as shown in Equations 13 and 14, which enables the particles to accumulate near the temperature inversions.

We find that a fit to the observed aerosol concentration can be obtained regardless of the bulk radius of the aggregates injected, as the fractal dimension $D_f = 2$ (see Eq. 14 and discussion thereafter). We thus choose the retrieved radius of 1.28 $\mu m$ from Fan et al. (2022). It is worth noting that coagulation does not strongly impact the distribution of these aggregates due to their low number density. For $N_i = N_j = 10^3$ and a number density of $1\,\text{cm}^{-3}$, the reaction rate is approximately $10^{-11}\text{cm}^{-3}\text{s}^{-1}$. The characteristic time scale of the coagulation to significantly change the distribution is then $10^{11}$ seconds, which is more than 10 Pluto year cycles. This timescale is much longer than the characteristic timescale of the build-up of the atmosphere, so the interactions among the aggregates themselves are not important. Furthermore, a large population of smaller aggregates is not observed, ruling out the path in which the aggregates can be generated while airborne through coagulation. Therefore, the aggregates could only come from being lofted from the surface. We speculate that it may be possible for the aggregates to form from sintering of individual spherical haze particles once they have settled onto the surface due to actions of the surface ice matrix, but laboratory measurements will be needed to test this idea.

What happens to the lofted aerosols as the atmosphere continues to build-up? The shape of the vertical distribution of aerosols comes from the balance between the settling speed of the aerosols and the growth of the atmosphere. During the build-up of the atmosphere, if the upward speed of the atmosphere exceeds the settling speed of the aerosols, the aerosol will effectively move up in altitude. As the atmosphere keeps building up, the aerosols that enter the atmosphere later move up faster as the settling speed in the denser lower atmosphere is slower. This results in an accumulation effect, creating a layer of aerosols with a number density higher than that at the surface.

This story ends when the atmosphere starts to decline. The distribution of the aggregates at several other years are shown in Figure 5. We note that inversions only exist at the late phase of atmospheric growth. We predict that, as the atmosphere starts to decline, due to the lack of continued supply, the large aggregate mode of the aerosol will quickly disappear.

**Discussions and Conclusions**

In this work, we have shown that the sublimation flow of the atmosphere could play an important role in forming the bimodal distribution of aerosols observed on Pluto. The KINAERO model successfully reproduces the number density inversion of aggregates at about 20 km altitude, and a surface number density of 0.139 cm$^{-2}$ of the aggregates is suggested in the sublimation flow. We generate the smaller spherical aerosol mode through photochemistry and coagulation, reproducing the spherical particle case from Gao et al., (2017). While the spherical mode will continue to be produced photochemically as Pluto continues in its orbit, though likely at a reduced rate as the atmosphere begins to collapse (Figure 2), the aggregate mode will disappear when the sublimation flow is halted. Hence, we claim that the observed bimodal distribution of haze particles is a transient and dynamic phenomenon, which can be tested by future JWST observations, or even missions that revisits Pluto.

We have made several simplifying approximations in this study. For example, due to the change of pressure in the atmosphere and the injection of chemical species from the sublimation flow, simulating the state-of-the-art complex chemistry network will take a prohibitively long time. Also, the seasonal cycle calculations are not scalable. Hence, we used a simpler photochemistry model that only simulates up to C2. Different time-stepping solutions would be required if more complex chemistry is desired.

Another possible improvement to the current work is to consider 3D atmospheric dynamics. As the most significant source and sink of the atmosphere are in Sputnik Planitia, the dynamical effects would be important, and such spatial asymmetries can strongly affect the observations (e.g. Chen et al., 2021). We suggest that future works use 2D or 3D models to better resolve the effects of a sublimation plume originating from Sputnik Planitia.

Two crucial processes remain poorly understood and have yet to be accurately modeled. The first involves the formation of aggregate structures on Pluto's surface, while the second pertains to the conditions required for aerosol particle entrainment in the sublimation flow. Given the complexities, these processes cannot be effectively modeled through analytical or numerical methods, leading to our parameterized approach. Future experimental works that investigate aerosol-surface ice interactions and modeling studies that couple haze particles to atmospheric dynamics on orbital time scales would be needed to shed light on these problems.

Finally, our model necessitates that photochemistry and coagulation are not a major source of aggregates in the near-surface atmosphere, which differs from conclusions of previous haze models (Gao et al. 2017; Lavvas et al. 2021). Condensation of ices on the sedimenting aerosol particles could result in the collapse of aggregates and an increase in $D_f$ (Lavvas et al. 2021), but that could result in larger spherical particles than the observed small spherical mode due to the added mass of condensed material. Another possibility is that particle charging is much more intense, which would reduce the coagulation efficiency and particle size of the falling haze (Lavvas et al. 2010; Gao et al. 2017). Upper atmosphere/ionosphere modeling would be needed to quantify the particle radius-to-charge ratio of haze particles in Pluto's atmosphere.

**Figures**

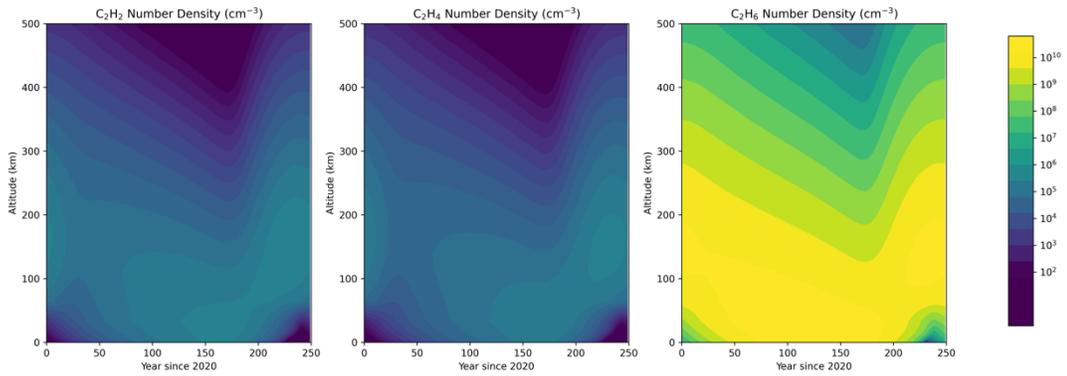

Figure 1. The base-10 logarithm of the number density in cm$^{-3}$ for the C2 species over a Pluto year. The dark regions near the surface at the beginning and end of the cycles mark the time when the atmosphere grows and fills the lower atmosphere with $CH_4$ and $N_2$, pushing the C2 species upwards and reducing their number densities near the surface.

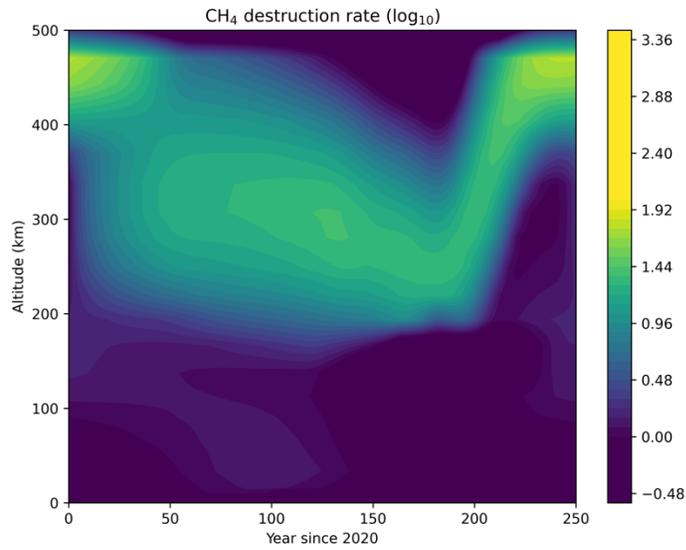

Figure 2. The destruction rate distribution of methane, in $\log_{10}$ under the unit of cm$^{-3}$s$^{-1}$.

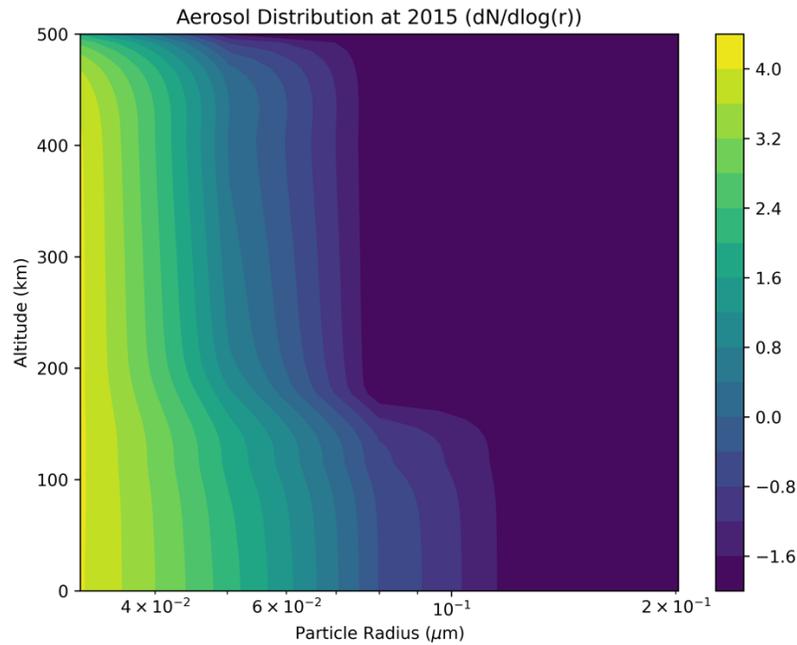

Figure 3. The number density distribution of the spherical haze mode, as produced by photochemistry and coagulation, in $\log_{10}$ under the unit of cm$^{-3}$.

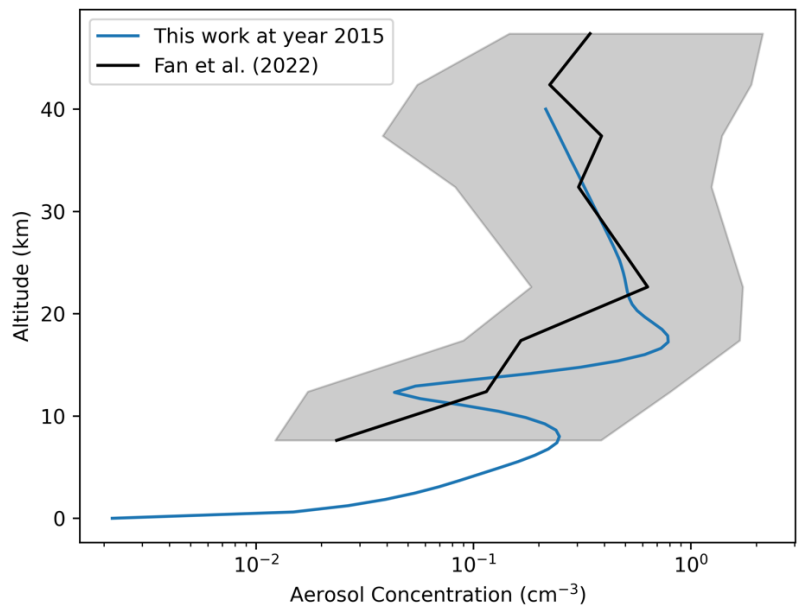

Figure 4. The distribution of aggregates with monomer size $r_0 = 10^{-2} \mu m$, aggregate size $R_a = 1.28\ \mu m$ in the year 2015, coinciding with the New Horizons flyby. The result of this work is compared with the retrieved profile (with uncertainty range in gray) of the aggregates from Fan et al. (2022).

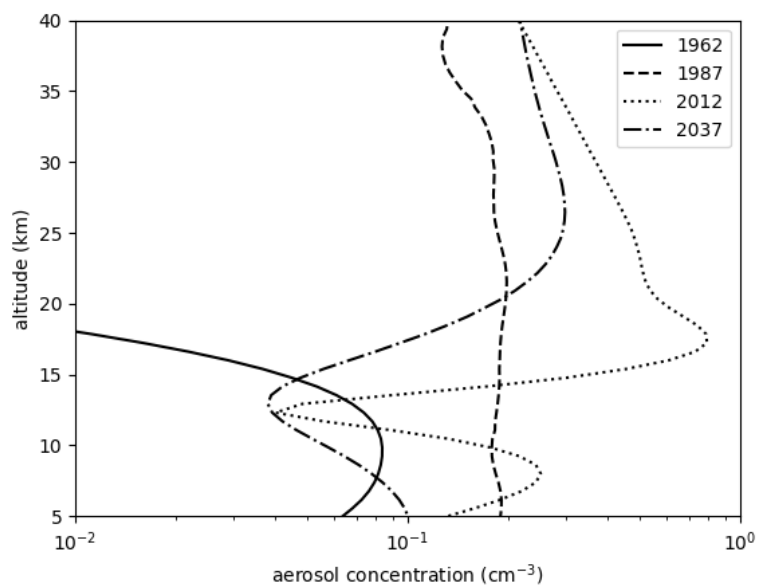

Figure 5. Number density distribution of the aggregates at different years. Note that the high concentration at high altitudes in year 2037 is not physically real, as we took a zero-gradient top boundary condition.

**Acknowledgement**

This research was supported in part by NASA's New Horizons mission to the Pluto system under the grant NASA NFDAP 80NSSC19K0823.

**References**

Adams, D., Luo, Y., & Yung, Y. (2022). Hydrocarbon chemistry in the atmosphere of a Warmer Exo-Titan. *Frontiers in Astronomy and Space Sciences*, *9*, 823227. https://doi.org/10.3389/fspas.2022.823227

Allen, M., Yung, Y. L., & Waters, J. W. (1981). Vertical transport and photochemistry in the terrestrial mesosphere and lower thermosphere (50–120 km). *Journal of Geophysical Research: Space Physics*, *86*(A5), 3617–3627. https://doi.org/10.1029/JA086iA05p03617

Bertrand, T., & Forget, F. (2016). Observed glacier and volatile distribution on Pluto from atmosphere–topography processes. *Nature*, *540*(7631), Article 7631. https://doi.org/10.1038/nature19337

Bertrand, T., Forget, F., White, O., Schmitt, B., Stern, S. A., Weaver, H. A., Young, L. A., Ennico, K., Olkin, C. B., & Team, the N. H. S. (2020). Pluto's Beating Heart Regulates the Atmospheric Circulation: Results From High-Resolution and Multiyear Numerical Climate Simulations. *Journal of Geophysical Research: Planets*, *125*(2), e2019JE006120. https://doi.org/10.1029/2019JE006120

Chen, S., Young, E. F., Young, L. A., Bertrand, T., Forget, F., & Yung, Y. L. (2021). Global climate model occultation lightcurves tested by August 2018 ground-based stellar occultation. *Icarus*, *356*, 113976.

Cheng, A. F., Summers, M. E., Gladstone, G. R., Strobel, D. F., Young, L. A., Lavvas, P., Kammer, J. A., Lisse, C. M., Parker, A. H., Young, E. F., Stern, S. A., Weaver, H. A., Olkin, C. B., & Ennico, K. (2017). Haze in Pluto's atmosphere. *Icarus*, *290*, 112–133. https://doi.org/10.1016/j.icarus.2017.02.024

Fan, S., Gao, P., Zhang, X., Adams, D. J., Kutsop, N. W., Bierson, C. J., Liu, C., Yang, J., Young, L. A., Cheng, A. F., & Yung, Y. L. (2022). A bimodal distribution of haze in Pluto's atmosphere. *Nature Communications*, *13*(1), Article 1. https://doi.org/10.1038/s41467-021-27811-6

Gao, P., Fan, S., Wong, M. L., Liang, M.-C., Shia, R.-L., Kammer, J. A., Yung, Y. L., Summers, M. E., Gladstone, G. R., Young, L. A., Olkin, C. B., Ennico, K., Weaver, H. A., & Stern, S. A. (2017). Constraints on the microphysics of Pluto's photochemical haze from New

Horizons observations. *Icarus*, *287*, 116–123.

https://doi.org/10.1016/j.icarus.2016.09.030

Gladstone, G. R., Stern, S. A., Ennico, K., Olkin, C. B., Weaver, H. A., Young, L. A., Summers, M. E., Strobel, D. F., Hinson, D. P., Kammer, J. A., Parker, A. H., Steffl, A. J., Linscott, I. R., Parker, J. Wm., Cheng, A. F., Slater, D. C., Versteeg, M. H., Greathouse, T. K., Retherford, K. D., … THE NEW HORIZONS SCIENCE TEAM. (2016). The atmosphere of Pluto as observed by New Horizons. *Science*, *351*(6279), aad8866. https://doi.org/10.1126/science.aad8866

Grundy, W. M., Bertrand, T., Binzel, R. P., Buie, M. W., Buratti, B. J., Cheng, A. F., Cook, J. C., Cruikshank, D. P., Devins, S. L., Dalle Ore, C. M., Earle, A. M., Ennico, K., Forget, F., Gao, P., Gladstone, G. R., Howett, C. J. A., Jennings, D. E., Kammer, J. A., Lauer, T. R., … Zhang, X. (2018). Pluto's haze as a surface material. *Icarus*, *314*, 232–245. https://doi.org/10.1016/j.icarus.2018.05.019

Johnson, P. E., Young, L. A., Protopapa, S., Schmitt, B., Gabasova, L. R., Lewis, B. L., Stansberry, J. A., Mandt, K. E., & White, O. L. (2021). Modeling Pluto's minimum pressure: Implications for haze production. *Icarus*, *356*, 114070. https://doi.org/10.1016/j.icarus.2020.114070

Kasting, J. F., Zahnle, K. J., & Walker, J. C. G. (1983). Photochemistry of methane in the Earth's early atmosphere. *Precambrian Research*, *20*(2), 121–148. https://doi.org/10.1016/0301-9268(83)90069-4

Krasnopolsky, V. A., & Cruikshank, D. P. (1999). Photochemistry of Pluto's atmosphere and ionosphere near perihelion. *Journal of Geophysical Research: Planets*, *104*(E9), 21979–21996. https://doi.org/10.1029/1999JE001038

Kutsop, N. W., Hayes, A. G., Buratti, B. J., Corlies, P. M., Ennico, K., Fan, S., Gladstone, R., Helfenstein, P., Hofgartner, J. D., Hicks, M., Lemmon, M., Lunine, J. I., Moore, J., Olkin, C. B., Parker, A. H., Stern, S. A., Weaver, H. A., Young, L. A., & Team, T. N. H. S. (2021). Pluto's Haze Abundance and Size Distribution from Limb Scatter Observations by MVIC. *The Planetary Science Journal*, *2*(3), 91. https://doi.org/10.3847/PSJ/abdcaf

Lavvas, P., Lellouch, E., Strobel, D. F., Gurwell, M. A., Cheng, A. F., Young, L. A., & Gladstone, G. R. (2021). A major ice component in Pluto's haze. *Nature Astronomy*, *5*(3), Article 3. https://doi.org/10.1038/s41550-020-01270-3

Lavvas, P., Yelle, R. V., & Griffith, C. A. (2010). Titan's vertical aerosol structure at the Huygens landing site: Constraints on particle size, density, charge, and refractive index. *Icarus*, *210*(2), 832–842. https://doi.org/10.1016/j.icarus.2010.07.025

Stern, S. A., Bagenal, F., Ennico, K., Gladstone, G. R., Grundy, W. M., McKinnon, W. B., Moore, J. M., Olkin, C. B., Spencer, J. R., Weaver, H. A., Young, L. A., Andert, T., Andrews, J., Banks, M., Bauer, B., Bauman, J., Barnouin, O. S., Bedini, P., Beisser, K., … Zirnstein, E. (2015). The Pluto system: Initial results from its exploration by New Horizons. *Science*, *350*(6258), aad1815. https://doi.org/10.1126/science.aad1815

Wong, M. L., Fan, S., Gao, P., Liang, M.-C., Shia, R.-L., Yung, Y. L., Kammer, J. A., Summers, M. E., Gladstone, G. R., Young, L. A., Olkin, C. B., Ennico, K., Weaver, H. A., & Stern, S. A. (2017). The photochemistry of Pluto's atmosphere as illuminated by New Horizons. *Icarus*, *287*, 110–115. https://doi.org/10.1016/j.icarus.2016.09.028

Young, L. A., Kammer, J. A., Steffl, A. J., Gladstone, G. R., Summers, M. E., Strobel, D. F., Hinson, D. P., Stern, S. A., Weaver, H. A., Olkin, C. B., Ennico, K., McComas, D. J., Cheng, A. F., Gao, P., Lavvas, P., Linscott, I. R., Wong, M. L., Yung, Y. L., Cunningham,

N., … Versteeg, M. (2018). Structure and composition of Pluto's atmosphere from the New Horizons solar ultraviolet occultation. *Icarus*, *300*, 174–199. https://doi.org/10.1016/j.icarus.2017.09.006

Zhang, X., Strobel, D. F., & Imanaka, H. (2017). Haze heats Pluto's atmosphere yet explains its cold temperature. *Nature*, *551*(7680), 352–355. https://doi.org/10.1038/nature24465